\documentstyle[aps,preprint,eqsecnum]{revtex}  
\begin{document}                
\preprint{UOS-TP-97-1005}
\title{Boundary degrees of freedom in fractional quantum Hall
effect: Excitations on common boundary of two samples}

\author{Dongsu Bak$^a$\footnote{E-mail: dsbak@mach.scu.ac.kr},
Sang Pyo Kim$^b$\footnote{E-mail: sangkim@knusun1.kunsan.ac.kr},
Sung Ku Kim$^c$\footnote{E-mail: skkim@theory.ewha.ac.kr},
Kwang-Sup Soh$^d$\footnote{E-mail: kssoh@phya.snu.ac.kr},
Jae Hyung Yee$^e$\footnote{E-mail: jhyee@phya.yonsei.ac.kr}}

\address{$^a$Department of Physics, University of Seoul,
Seoul 130-743,   Korea\\
$^b$Department of Physics, Kunsan National
University, Kunsan 573-701, Korea\\
$^c$Department of Physics, Ewha Women's University,
Seoul 120-750, Korea\\
$^d$Department of Physics Education, Seoul National University,
Seoul 151-742, Korea\\
$^e$Department of Physics, Yonsei University, Seoul 120-749, Korea}

\maketitle
\begin{abstract}                
Using the Carlip's method 
we have derived the boundary action for the fermion
Chern-Simons theory of quantum Hall effects on a planar region 
with a boundary.
We have computed both the bulk and edge responses of currents 
to the external
electric field.
From this we obtain the
well-known anomaly relation and the boundary Hall current without
introducing any {\it ad hoc} assumptions such as the chirality 
condition. 
In addition, the edge current on the common
boundary of two samples is found to be proportional to 
the difference
between Chern-Simons coupling strengths.
\end{abstract}
\vspace{1cm}

\newpage

\section{Introduction}
The phenomenological theory of fractional quantum Hall 
effect (FQHE)
based on the Chern-Simons gauge field coupled to 
interacting fermions was
advanced first by Lopez and Fradkin\cite{Lopez}, and 
by Halperin and
coworkers \cite{Halperin}, and has been remarkably successful  
in explaining
the fractional Hall conductance. Recently the theory has been
reformulated in aesthetically satisfying manner by Gimm 
and Salk\cite{Gimm}.
This theory of FQHE consists of spinless (or completely
polarized) electron field coupled to both external 
electromagnetic and
Chern-Simons gauge fields, and is described by the 
action\cite{Gimm}
\begin{eqnarray}
S\!=\!\int\! d^3\! x \,\psi ^* \!\left(iD_0\! + 
\!\mu \!-\!
\frac{1}{2m} {\bf D}^2 \right)\psi\! +\!
S_{CS}[a]\! -\! \frac{1}{2} \!\int\! d^3\!x d^3\!x' 
(|\psi (x)|^2 \!\!-\!
\bar{\rho}) V(x,\!x')( |\psi(x')|^2 \!\!-\! \bar{\rho}),
\label{e11}
\end{eqnarray}
with
\begin{equation}
 S_{CS}[a]=\int d^3x\frac{\alpha}{2}
\epsilon ^{\mu \nu \rho}a_\mu\partial_\nu a_\rho\ ,
\label{e12}
\end{equation}
where $D_0=\partial_0 - ie(A_0 -a_0)$, ${\bf D} =
{\nabla}+ie({\bf A}-{\bf a})$, and
 $(\psi , A_{\mu}, a_{\mu})$ represent, respectively, the 
electron,
external electromagnetic and Chern-Simons gauge potentials. 
$\mu$ and $\bar{\rho}$
denote the chemical potential and the average 
electron density,
respectively, and $V$ is the pair potential between 
electrons. If we
drop the Chern-Simons action $S_{CS}$ from (\ref{e11}), 
then the action
describes the integral quantum Hall effect\cite{Fradkin}.

This fermion Chern-Simons theory of FQHE, however, 
does not deal with
the possible boundary excitations of 2-dimensional 
quantum Hall droplet.
The description of quantum Hall phenomena by 
Chern-Simons gauge theory
was in fact initiated by the observation that 
the essential features of
both a droplet of incompressible 2-dimensional electron gas in the
quantum Hall regime and a pure Chern-Simons theory in a bounded
2-dimensional surface, are determined by the edge 
excitations at the
boundary\cite{Witten}. Wen\cite{Wen} used this 
observation to deduce
that the edge states form a representation of Kac-Moody 
algebra, which
indicates the existence of physical observable at 
the boundary, and
Stone\cite{Stone} and Balachandran, {\it et al.}\cite{Balachandran} 
have
subsequently shown that the representation is 
isomorphic to that
generated by a chiral scalar field theory on 
the boundary. By requiring
the gauge invariance at the boundary they derived 
the boundary action
which consists of a scalar field chirally coupled 
to a gauge field. This
has led to the (1+1)-dimensional effective field theory
approach\cite{Chandar}
that accounts for the quantum Hall effects mainly as edge
phenomena.

Recently Carlip\cite{Carlip} had observed that the 
Chern-Simons action
(\ref{e12}) defined on a bounded surface cannot have 
classical extrema
due to the boundary contribution, and that
the
variational principle does require a modification of 
the action (\ref{e12})
by a boundary action.
He used this observation to derive a boundary action
for the  Chern-Simons gravity in (2+1)-dimensions.
Especially it was shown that the  gauge variance of 
the boundary action
makes the would-be gauge degrees of
freedom dynamical.

The purpose of this paper is to study the boundary 
excitations of
a quantum Hall droplet starting from the well-established
fermionic Chern-Simons model of FQHE, with help 
of the  Carlip's
method.
We start from the action (\ref{e11}) and
use only the standard procedures of quantum field theory
without introducing  any {\it ad hoc}
assumptions such as the chirality condition.
We then proceed to clarify  the resulting edge excitations of
boundaries.   Especially we wish to
consider two samples
joined together with a common boundary. However,
the complicated contact interactions
may arise  if two quantum Hall samples
differing in their electronic properties, are
merged. To avoid such unnecessary complexity,
we, instead, choose to consider a boundary formed
by a stepwise external magnetic field within a sample. It 
turns out that
this  effectively generates two regions with distinct
Chern-Simons coupling strengths.
Subsequently, we analyze in detail
the effects of the  boundary to the bulk,
the boundary excitations, and their experimental implications.
Should the  results be confirmed
by experiments, it would mean that the theory (\ref{e11}) 
indeed
provides  a good
field theoretic description of the quantum Hall effect.
electron system.

In the next section we start from the Chern-Simons 
action (\ref{e12})
and show how the boundary degrees of freedom arise both 
by the Carlip's
original method, and by the path integral approach. We 
then show that
the so called edge phenomena can be understood  from the
boundary action supplemented to the bulk part in (\ref{e12}).
In section III, we consider the case of two quantum
Hall droplets sharing a common boundary, study the edge 
excitations,
and
derive the boundary current.
In the last section we
conclude with  discussions on our results.

\section{Boundary degrees of freedom in Chern-Simons 
gauge theory}
In this section we consider the pure Chern-Simons 
gauge theory, and show 
how the gauge variance of the theory at the boundary 
gives rise to the 
physical degrees of freedom by the Carlip's original 
method\cite{Carlip}
and also by the path integral approach. We then consider 
the theory with 
an external source and study the effect of these 
boundary degrees of freedom on the
quantum Hall phenomena.

\subsection{Carlip's method}
The variation of the Chern-Simons action (\ref{e12}) defined 
on a three-manifold
M is 
\begin{equation}
\delta S_{CS} = \frac{\alpha}{2} \int _M d^3x \partial _{\nu}
[\epsilon^{\mu \nu \rho} a_\mu \delta a_{\rho}] + \alpha \int _M
d^3x \epsilon^{\mu \nu \rho} \delta a_{\mu} \partial _{\nu} a_{\rho}.
\label{e21}
\end{equation}
The first term of Eq.(\ref{e21}) is the boundary term 
which breaks the gauge
invariance. This action has no classical extrema for 
generic variations
on the boundary.
In order for the theory to admit classical solutions, 
therefore,
$S_{CS}$ must be supplemented by a surface term so that its
variation exactly cancels the first term of Eq.(\ref{e21});

\begin{equation}
\delta S_{\partial M} = - \frac{\alpha}{2} \int_M d^3x
\partial_{\nu}[\epsilon^{\mu \nu \rho} a_{\mu} \delta a_{\rho}].
\label{e22}
\end{equation}
The exact form of this boundary action depends on the 
choice of boundary
conditions.

For simplicity we take the spatial part of $M$ to be a 
half-plane
$(y<0)$ with the boundary given by $y=0$. 
 $x\rightarrow \theta$ and
On this manifold the variation of boundary action becomes 

\begin{equation}
\delta S_{\partial M} = - \frac{\alpha}{2} \int_{\partial M} 
dt dx(a_0 \delta a_1 -
a_1 \delta a_0).
\label{e23}
\end{equation}
If the field component $(a_0-a_1)$ is fixed at the 
boundary 

\begin{equation}
\delta (a_0-a_1)=0,
\label{e24}
\end{equation}
then the boundary action is given by 

\begin{equation}
S_{\partial M} = - \frac{\alpha}{4} \int _{\partial M} dt dx
(a_0 - a_1)(a_0 + a_1).
\label{e25}
\end{equation}
Although the total action now gives meaningful classical 
equations of
motion on the manifold $M+ \partial M$, it is still not 
gauge invariant.
One can make this gauge variance explicit by decomposing 
the gauge field
as

\begin{equation}
a_{\mu} = \tilde{a}_\mu + \partial _{\mu} \Lambda,
\label{e26}
\end{equation}
where $\tilde{a}_\mu$ is a gauge fixed potential. Then the 
total action
becomes

\begin{eqnarray}
S_{M+\partial M}[a]&=&\frac{\alpha}{2}\int_M d^3 x
\epsilon^{\mu\nu\rho}\tilde{a}_{\mu}\partial_{\nu}\tilde{a}_{\rho}+
\frac{\alpha}{4}\int_{\partial
M} dtdx(\tilde{a}_0-\tilde{a}_1)(\tilde{a}_0+\tilde{a}_1) 
\nonumber \\
&&+\frac{\alpha}{4}\int_{\partial M}dtdx[\partial_a \Lambda\partial^a
\Lambda+2(\tilde{a}_0-\tilde{a}_1)(\partial_0 \Lambda+
\partial_1 \Lambda)]
\label{e27}
\end{eqnarray}
where the latin index $a$ denotes the space-time index on 
the boundary,
{\it i.e.}, $a=0,1$, $(\partial ^a = (\partial_0, 
-\partial_1))$ 
while $\mu,\nu,\rho = 0,1,2$. This action explicitly
shows that $\Lambda (x)$ became a dynamical variable 
with $\partial_1$ legitimate kinetic term.
The would-be gauge variable $\Lambda (x)$ has thus 
become a dynamical field  
on $\partial M$. This boundary action can be recognized 
as a chiral
Wess-Zumino-Witten action\cite{Wess}.

\subsection{Path integral approach}
Although we have derived the boundary action(\ref{e27}) 
by requiring
the existence of classical solutions on a manifold with 
a boundary, one 
might wonder why the Chern-Simons theory, which has
no physical degrees of freedom, gives rise to a theory 
with dynamical
degrees of freedom at the boundary. The origin of this 
dynamical degrees
of freedom is the gauge non-invariance of the theory 
at the boundary.
This fact can be better seen in the following path integral 
approach.

The generating functional or partition function for the 
Chern-Simons
theory on a manifold $M+\partial M$ is defined by

\begin{equation}
Z=\int D a_{\mu} e^{i\{S_{CS}[a]+S_{\partial M}[a]\}}
\label{e28}
\end{equation}
where $S_{CS}$ is given by Eq.(\ref{e12}) and 
$S_{\partial M}$ by
Eq.(\ref{e25}). 
Following the Faddeev-Popov procedure, we introduce

\begin{equation}
\Delta(a) \equiv \int D\Lambda~\delta[F(a_{\Lambda})-C(x)],
\label{e29}
\end{equation}
where $\Lambda(x)$ is a gauge parameter and $F(a)$ is 
a gauge fixing function.
We use the covariant gauge $F(a)=\partial _{\mu}a^{\mu}$ 
for simplicity.
$a_{\Lambda}$ in Eq.(\ref{e29}) is a gauge transformed 
potential 
$(a_{\Lambda})_\mu = a_{\mu}-\partial_{\mu}\Lambda$. 
By using the change
of variable method one can easily show that

\begin{equation}
\Delta(a)=[\det \frac{\delta F(a_\Lambda)}
{\delta \Lambda}]^{-1}=[\det
\partial_{\mu}\partial^{\mu}]^{-1}=\Delta(a_{\Lambda}),
\label{e210}
\end{equation}
and that $\Delta(a)$ is gauge invariant and 
independent of function
$C(x)$.

Substituting Eq.(\ref{e210}) into Eq.(\ref{e28}) 
we have 

\begin{eqnarray}
Z&=&\int D a_{\mu} e^{i\{S_{CS}[a]+S_{\partial
M}[a]\}}\Delta^{-1}(a)\Delta(a) \nonumber \\
&=&[\det(\partial_{\mu}\partial^{\mu})] \int D
\Lambda D a_{\mu}
e^{i\{S_{CS}[a]+S_{\partial M}[a]\}}
\delta[F(a_{\Lambda})-C].
\label{e211}
\end{eqnarray}
By averaging over the function $C(x)$ and making 
gauge transformation of
$a_{\mu}$ so that $(a_{\Lambda})_\mu$ transforms 
back to $a_\mu$, we obtain

\begin{eqnarray}
Z &=& \int D C D\Lambda Da_{\mu} e^{i \int C(x)^2 /2 \beta ~d^3x} 
\delta [F(a_\Lambda )-C]
e^{i\{S_{CS}[a]+S_{\partial M}[a]\}} \nonumber \\
&=&\int D C D\Lambda Da_{\mu} e^{i\int C^2/2\beta~d^3x} 
\delta[F(a_\Lambda)-C]
e^{i\{S_{CS}[a_{\mu}+\partial_{\mu}\Lambda]+S_{\partial M}
[a_{\mu}+\partial_{\mu}\Lambda]\}} \nonumber \\
&=&N\int D\Lambda Da_{\mu} e^{i\{S_{CS}[a_{\mu}+
\partial_{\mu}\Lambda]+S_{\partial M}[a_{\mu}+
\partial_{\mu}\Lambda]+\frac{1}{2\beta}(\partial_{\mu}a^{\mu})^2 \}}
\label{e212}
\end{eqnarray}
where $\beta$ is a gauge fixing parameter and the 
use has been made of the
fact that $Z$ is independent of the function $C(x)$, 
and $S_{CS}$ plus
$S_{\partial M}$ is not gauge invariant. Note that 
the exponent in the
last line of Eq.(\ref{e212}) is exactly the Carlip's 
action (\ref{e27}) in the
covariant gauge. 

If the classical action $S_{CS}+S_{\partial M}$ were 
gauge invariant,
then $\Lambda$-integration in Eq.(\ref{e212}) would 
be trivial giving 
an infinite
normalization constant, and the above procedure would 
be an ordinary way
of fixing the gauge. But for the Chern-Simons theory 
on a manifold with
a boundary, the classical action is not gauge 
invariant at the boundary,
and the $\Lambda(x)$-integration becomes non-trivial. 
In fact
$\Lambda$-variable becomes a dynamical variable with 
a bona fide kinetic
term as can be seen from Eq.(\ref{e27}). We thus see 
how the
would-be gauge variable becomes physical in this case.

\subsection{One parameter family of the boundary actions}
As pointed out by Carlip\cite{Carlip} the form of 
the boundary action
depends on the choice of the boundary conditions 
we impose at the
boundary. In deriving the boundary action (\ref{e27}) 
we have fixed the
potential $(a_0 - a_1)$ at the boundary by 
requiring Eq.(\ref{e24}). If we
fix $(a_0 -va_1)$ instead, 

\begin{equation}
\delta(a_0-va_1)=0,
\label{e213}
\end{equation}
at the boundary with a parameter $v$, then we have

\begin{equation}
\delta S_{CS}=\frac{\alpha}{4v}\int_{\partial M} dtdx
(a_0-va_1)(a_0+va_1),
\label{e214}
\end{equation}
and the boundary action becomes

\begin{equation}
S_{\partial M}=\frac{\alpha}{4v}\int_{\partial M} 
dtdx[\partial_a
\Lambda\partial^a
\Lambda+2(\tilde{a}_0-v\tilde{a}_1)(\partial_0\Lambda+
v\partial_1\Lambda)].
\label{e215}
\end{equation}
We thus see that the boundary action is dependent on 
the parameter $v$ which has
a dimension of velocity. We will discuss more on the role and
significance of this parameter later.

\subsection{The effect of boundary action on the 
quantum Hall phenomena}
Although we have the boundary action for the 
$\Lambda$-field in Eq.(\ref{e27}), we
do not yet know what the $\Lambda$-field represents, 
namely, whether it
describes dynamics of a physical particle, a 
quasiparticle, or it only
describes some intermediate stage of a physical process. 
We therefore
integrate out the $\Lambda$-field in the path integral 
method, and find
out what the effect of the $\Lambda$-field on the quantum 
Hall current is.

To do this we couple the theory to an external current 
$J^{\mu}(x)$, so
that the total action $S_T$ reads

\begin{eqnarray}
S_T&=&\int_M
d^3x \left(\frac{\alpha}{2}\epsilon^{\mu\nu\rho}a_{\mu}
\partial_{\nu}a_{\rho}
-a_{\mu}J^{\mu}\right)+\frac{\alpha}{4v}
\int_{\partial M}dtdx
(a_0-va_1)(a_0+va_1) \nonumber \\
&&+\frac{\alpha}{4v}\int_{\partial
M}dtdx[-\Lambda(\partial_0^2-v^2\partial_1^2-i\epsilon)
\Lambda+2(a_0-va_1)(\partial_0\Lambda+v\partial_1\Lambda)],
\label{e216}
\end{eqnarray}
where we have introduced the constant $v$ characterizing 
the choice of
the boundary conditions, and dropped the tilde over $a_{\mu}$ 
for notational
simplicity, but $a_{\mu}$ is understood to be the gauge 
fixed potential. Note
that we have introduced $i\epsilon$ prescription into 
the kinetic term of
$\Lambda$-field so that the functional integral over 
$\Lambda$ exists.
This defines the Green's function appearing in the 
effective action to
be Feynman Green's function, and we will discuss the 
physical
implications of this prescription later. Since the 
action (\ref{e216}) is
quadratic in ${\Lambda}$-field and we have introduced 
the Feynman
prescription, the path integral over ${\Lambda}$ can 
easily be performed;

\begin{eqnarray}
Z&=&\int Da_{\mu} D\Lambda e^{iS_T[a,\Lambda]} \nonumber \\
&=&N\int Da_{\mu} e^{iS_{eff}[a]},
\label{e217}
\end{eqnarray}
where

\begin{eqnarray}
S_{eff}[a]&=&\int_M
d^3x[\frac{\alpha}{2}\epsilon^{\mu\nu\rho}a_{\mu}
\partial_{\nu}a_{\rho}
-a_{\mu}J^{\mu}]+\frac{\alpha}{4v}\int_{\partial M}dtdx
(a_0-va_1)(a_0+va_1) \nonumber \\
&&+\frac{\alpha}{4v}\int_{\partial
M}dtdx(\partial_0+v\partial_1)(a_0-va_1)
\frac{1}{\partial_0^2-v^2\partial_1^2-i\epsilon}
(\partial_0+v\partial_1)(a_0-va_1).
\label{e218}
\end{eqnarray}
By varying $S_{eff}[a]$ with respect to the variation 
of $a_{\mu}$ and setting 
$\delta _a S_{eff}[a]=0$, we obtain
 
\begin{equation}
J^{\mu}(x)=\alpha\theta(-y)\epsilon^{\mu\nu\rho}
\partial_{\nu}a_{\rho}
+\frac{\alpha}{2v}\delta(y)m^{\mu}
\label{e219}
\end{equation}
where $m^{\mu} = (m, -vm, 0)$, $\theta(y)$ is the 
step function, and 

\begin{eqnarray}
m&=&(a_0+va_1)-(\partial_0+v\partial_1)\frac{1}
{\partial_0^2-v^2\partial_1^2-i\epsilon}
(\partial_0+v\partial_1)(a_0-va_1) \nonumber \\
&=&\frac{1}{\partial_0^2-v^2\partial_1^2-i\epsilon}
(\partial_0+v\partial_1)2v\epsilon^{ab}
\partial_a a_b \nonumber \\
&=&2v\frac{1}{\partial_0^2-v^2\partial_1^2-i\epsilon}
(\partial_0+v\partial_1)E_x.
\label{e220}
\end{eqnarray}
In Eq.(\ref{e220}) we have used the fact that the 
$x$-component of electric field is
defined by $E_x=\epsilon^{ab}\partial_a a_b$. Note 
that the first term of the 
first line of Eq.(\ref{e220}) comes from the bulk 
contribution, namely, the first term 
of Eq.(\ref{e218}).

If we were interested in the integral quantum Hall 
effect, we would drop the
Chern-Simons action (\ref{e12}) from Eq.(\ref{e11}), 
integrate over the fermion
field $\psi$ up to the one-loop order, and would 
obtain the effective
action in the form of Eq.(\ref{e218}) without external current
where $a_{\mu}$ now represents the fluctuation of external
electromagnetic field\cite{Fradkin}. Then the Hall 
current would be
determined by $\frac{\delta S_{eff}}{\delta a_{\mu}}$, 
which would be exactly 
the current $J^{\mu}(x)$ given in Eq.(\ref{e219}). 
Therefore Eq.(\ref{e219}) determines
the structure of the integral Hall current on a manifold 
with boundary.

If we look at the current in 3-dimensional manifold 
$M+\partial M$, the current 
$J^{\mu}$ of Eq.(\ref{e219}) is conserved:

\begin{equation}
\partial_{\mu}J^{\mu}=0.
\label{e221}
\end{equation}
However, if we consider what happens only at the 
boundary $\partial M$, 
the current is not conserved. This is the realization 
of the well-known
Callan-Harvey anomaly cancellation mechanism\cite{Callan}. 
The current
at the boundary is

\begin{equation}
J_B^a=\frac{\alpha}{2v}m^a,
\label{e222}
\end{equation}
and the divergence of this current is 

\begin{equation}
\partial_a J_B^a=\alpha \epsilon^{ab} \partial_a a_b = 
\alpha E_x,
\label{e223}
\end{equation}
which is the current anomaly relation discussed in detail 
by the authors
of Refs.\cite{Stone,Balachandran,Chandar}, and explains 
the integral
quantum Hall effect with $\alpha$ being the Hall conductivity. 
This shows that
the integral quantum Hall effect can be described as an 
edge phenomenon.
Note that the anomaly relation is independent of the 
parameter $v$ which 
represents the choice of boundary conditions. 
As noted above a part of the anomalous edge current
$J^a$(through $m^a$) comes from
the bulk action, the first term of Eq.(\ref{e216}). 
This fact has been
discussed by authors of Refs.\cite{Chandar,Naculich}. 
It is amusing to
note that in our approach this phenomenon can be 
understood from
the first principle once we start from the 
phenomenological action
(\ref{e11})

\section{Fractional quantum Hall effect on the 
common boundary of two samples}
In the last section we have shown that the 
Carlip's method of finding
the boundary action applied to fermion Chern-Simons
 theory explains 
the important aspects of the integral quantum 
Hall phenomena by
the field theoretic method without further 
{\it ad hoc}
assumptions. In this section we apply the same method 
to the
fermion Chern-Simons theory of FQHE on a manifold $M$, 
which consists of
two submanifolds $M_1$ and $M_2$ sharing a common 
boundary $\partial
M$. The spatial parts of $M_1$ and $M_2$ consist 
of the lower $(y<0)$ and
the upper $(y>0)$ half-planes, respectively, and 
the boundary is defined by
the equation, $y=0$. In $M_1$ and $M_2$, we assume that 
electronic properties are the same except that 
electrons couple
to Chern-Simons gauge fields with the coupling constants 
$\alpha_1$ and
$\alpha_2$, respectively.

This system is described by the action (\ref{e11}) with 
the Chern-Simons
term replaced by

\begin{eqnarray}
S_{CS}[a,\Lambda] &=& \frac{1}{2} \int_{M_1+M_2}d^3x
[\theta(-y){\alpha_1}
+\theta(y)\alpha_2]\epsilon^{\mu\nu\rho}a_\mu 
\partial_\nu a_\rho 
\nonumber
\\ &&+ \frac{\alpha_1-\alpha_2}{4}\int_{\partial M}
dtdx[(a_0 -va_1)(a_0 +va_1)
\nonumber
\\ &&-\Lambda(\partial_0^2 -v^2 \partial_1^2)\Lambda +
  2(a_0-va_1)(a_0+v\partial_1)\Lambda].
\label{e31}
\end{eqnarray}
Note that the fermion part of the action Eq.(\ref{e11}) 
is not affected by the
presence of the boundary $\partial M$.

The Hubbard-Stratonovich transformation of the quartic 
interaction term
and the functional integration  over the fermion field 
yield the
effective action\cite{Lopez,Gimm},

\begin{equation}
S_{eff}= -iTr \ln(iD_0 + \mu + \lambda -\frac{1}{2m} 
{\bf D}^2) +
S_{CS}[a_\mu-\delta_{\mu 0}\lambda] + S[\lambda],
\label{e32}
\end{equation}
where

\begin{equation}
S[\lambda]= -\int_M d^3x \lambda (x) \bar{\rho} + 
\frac{1}{2} \int_M
d^3xd^3x' \lambda(x)V^{-1}(x-x') \lambda (x').
\label{e33}
\end{equation}
The Chern-Simons part of the action becomes, after 
the functional 
integration over $\Lambda$-field,   

\begin{eqnarray}
S_{CS}[a]&=& \frac{1}{2}
\int_{M_1+M_2}d^3x[\theta(-y){\alpha_1}
+\theta(y)\alpha_2]\epsilon^{\mu\nu\rho}a_\mu 
\partial_\nu a_\rho
+ \frac{\alpha_1-\alpha_2}{4v}\int_{\partial M}dtdx[(a_0 -va_1)(a_0
+va_1)
\nonumber
\\
&&+ (\partial_0+v\partial_1)(a_0-va_1)\frac{1}
{\partial_v^2-i\epsilon}
(\partial_0+v\partial_1)(a_0-va_1)],
\label{e34}
\end{eqnarray}
where $\partial _v^2 =\partial_0^2 - v^2\partial_1^2$.

The classical configurations $\bar{a}_\mu=
\left<a_{\mu}(x)\right>$ and
$\bar{\lambda}=\left<\lambda (x)\right>$ are determined by 
\begin{equation}
\left. \frac{\delta S_{eff}}{\delta a_\mu} 
\right|_{\bar{a},\bar{\lambda}} = 0
~~{\rm and}~~             
\left. \frac{\delta S_{eff}}{\delta \lambda} 
\right|_{\bar{a},\bar{\lambda}} = 0.
\label{e35}
\end{equation}
Following Gimm and Salk\cite{Gimm} we introduce 
\begin{equation}
A^{eff}_\mu = A_\mu -a_\mu,
\label{e36}
\end{equation}
so that
\begin{equation}
{\bf B}^{eff} = \nabla \times {\bf A}^{eff} =
\partial_1 A^2_{eff} - \partial_2 A^1_{eff}
\label{e37}
\end{equation}
describes the residual magnetic field after 
screened out by the
Chern-Simons magnetic field. Then Eq.(\ref{e35}) gives

\begin{equation}
\left<j^\mu\right>-[\alpha_1 \theta(-y) + \alpha_2 
\theta(y)] \epsilon
^{\mu\nu\rho}[\left<\partial_\nu A_\rho\right>-
\left<\partial_\nu A^{eff}_\rho\right>] 
-\frac{\alpha_1 -\alpha_2}{2v} m^\mu \delta(y) =0,
\label{e38}
\end{equation}

\begin{equation}
\left<j_0\right>-\bar{\rho}+\int_M d^3x'V^{-1}(x-x')
\left<\lambda(x')\right>=0,
\label{e39}
\end{equation}
where $m^{\mu}=(m,-vm,0)$,

\begin{eqnarray}
m&=&2v(\partial_0 + v\partial_1) \frac{1}{\partial_v^2-i\epsilon}
\epsilon^{ab}\partial_a\left<A_b-A_b^{eff}\right> 
\nonumber \\
&=&2v(\partial_0 + v\partial_1) \frac{1}
{\partial_v^2-i\epsilon}
\epsilon^{ab}\partial_a\bar{a}_b,
\label{e310}
\end{eqnarray}
and $j^{\mu}$ is the electromagnetic current of electrons.

Since $\left<j^0\right>$ is the average charge 
density of electrons, 
Eq.(\ref{e39}) leads to

\begin{equation}
\left<j^0\right>=\bar{\rho}, ~~ 
{\rm and} ~~\left<\lambda(x)\right>=0.
\label{e311}
\end{equation}
Then Eq.(\ref{e38}) implies, in the bulk 
regions $M_1$ and $M_2$, 

\begin{equation}
B_{eff}=B(y)-\left<b(y)\right>=B(y)-
\left( \frac{\theta(-y)}{\alpha_1} +
\frac{\theta(y)}{\alpha_2}\right) \bar{\rho},
\label{e312}
\end{equation}
where $B$ and $b$ denote the external and Chern-Simons 
magnetic fields, 
respectively, and the coupling constants
$\alpha_i$ are given by 

\begin{equation}
\frac{1}{\alpha_i} = 2m_i\phi_0 ~,~(i=1,2),
\label{e313}
\end{equation}
where $2m_i$'s are the numbers of Chern-Simons 
flux quantum $\phi_0$ in
each manifold $M_i$.

One possible setting to study the boundary effect 
of FQHE would be to join
two different
materials with different electronic properties with 
a common boundary.
Then the average electron density $\bar{\rho}$ would 
be in general a 
function of the
coordinate $y$. But this makes computations extremely 
complicated. Thus,
for computational simplicity, we will use the same 
material, 
but apply external magnetic fields differently in 
$M_1$ and $M_2$ 
so that the
Chern-Simons coupling constants become different. 
This makes the  study
of boundary effects very simple. This is
the reason why we have formulated the theory in 
such a way that the
average electron density is uniform over the whole 
manifold $M$ as
indicated by Eq.(\ref{e311}).

We thus apply an external magnetic field such that

\begin{equation}
B(y)=B_{eff}^0 +
\left(\frac{\theta(-y)}{\alpha_1}+ \frac{\theta(y)}
{\alpha_2}\right)
\bar{\rho}.
\label{e314}
\end{equation}
Then from Eq.(\ref{e312}) we find that the effective 
magnetic field 
felt by the electron is uniform over M:

\begin{equation}
B_{eff}=B_{eff}^0.
\label{e315}
\end{equation}
We can then use the results of Lopez and Fradkin
 \cite{Lopez} for computation of the effective 
action as a function of
the external field $A_{\mu}$. From Eq.(\ref{e314}) 
we find the fractional
filling factor,

\begin{equation}
\nu(y)=\theta(-y)\frac{1}{2m_1p+1}+\theta(y)\frac{1}{2m_2p+1},
\label{e316}
\end{equation}
where $p={N_e}/{N_{\phi}^{eff}}$ is the effective 
filling fraction with
$N_e$ and $N_{\phi}^{eff}$ denoting the total number 
of electrons and
the number of effective(residual) flux quanta, respectively. 
The effective Landau gap
is given by

\begin{equation}
\hbar\omega_c^{eff}=\frac{\hbar\omega_c}{1+
[\theta(-y)2m_1 + \theta(y)
2m_2]p}.
\label{e317}
\end{equation}

To obtain the linear response functions we again 
follow Ref.\cite{Gimm},
and separate the effective action Eq.(\ref{e32}) as 

\begin{equation}
S_{eff}=S_{eff}^{cl}+S^{(2)},
\label{e318}
\end{equation}
where $S_{eff}^{cl}$ is the effective action 
evaluated at the classical
configurations of the fields discussed above, 
and $S^{(2)}$ represents
the fluctuations around the classical solutions. 
Up to the quadratic
part in the fluctuation fields we have

\begin{eqnarray}
S^{(2)}&=&\frac{1}{2} \int_Md^3xd^3x' \delta 
A_\mu^{eff}(x) \Pi_{\mu\nu} 
(x,x')\delta A_\nu^{eff}(x') \nonumber \\
&&+ S_{CS}[\delta A_\mu - \delta A_\mu^{eff} - \delta
\lambda\delta_{\mu0}] + \frac{1}{2} \int_M d^3xd^3x' 
\delta\lambda(x)
V^{-1}(x-x')\delta\lambda(x')
\label{e319}
\end{eqnarray}
where $S_{CS}$ is given by Eq.(\ref{e34}), and 
$\Pi_{\mu\nu}$ is the
polarization tensor\cite{Lopez}.

After functional integration over $\delta \lambda (x)$, 
the fluctuation
part of the effective action becomes, up to the first order in
$\alpha_i$'s, 

\begin{eqnarray}
S_{eff}^{(2)}&=&\frac{1}{2} \int_Md^3xd^3x' \delta 
A_\mu^{eff} \Pi_{\mu\nu}
\delta A_\nu^{eff} + S_{CS}[\delta A_\mu-
\delta A_\mu^{eff}] \nonumber \\
&&+ \frac{1}{2}\int_Md^3x \frac{1}{\beta}(\partial_\mu \delta
A_{eff}^\mu)^2 + O(\alpha^2) \nonumber \\
&=& \frac{1}{2}\int_M d^3xd^3x'(\delta 
A_\mu-\delta a_\mu) [\Pi^{\mu\nu} -
\frac{1}{\beta}\partial^\mu \partial^\nu] 
(\delta A_\nu -\delta a_\nu) +
S_{CS}[a_\mu]+ O(\alpha ^2)
\label{e320}
\end{eqnarray}
where we have introduced the gauge fixing condition 
into the effective
action. Since we have evaluated $S_{eff}^{(2)}$ up to 
the first order in
$\alpha_i$'s, this result is valid when the coupling 
constants $\alpha_i$'s
 are small or when the numbers of attached flux quanta 
$2m_i$
are very large.

We now have to evaluate the functional integral over 
$\delta a_{\mu}$ to
 obtain the linear response functions. To do this we 
write Eq.(\ref{e320})
as

\begin{equation}
S_{eff}^{(2)} = \frac{1}{2} \int_Md^3xd^3x'
(\delta A_\mu-\delta a_\mu)
M^{\mu\nu}(x,x')(\delta A_\mu-\delta a_\mu) + 
\frac{1}{2} \int d^3xd^3x' \delta
a_\mu N^{\mu\nu}(x,x') \delta a_\nu + O(\alpha^2),
\label{e321}
\end{equation}
where $M^{\mu \nu} = \Pi^{\mu \nu}-\frac{1}{\beta} 
\partial ^{\mu}
\partial ^{\nu}$ and $N^{\mu\nu}$ is defined such 
that the second term of
Eq.(\ref{e321}) is the same as Eq.(\ref{e34}). 
It is convenient to 
write Eq.(\ref{e321}) in momentum space as 

\begin{eqnarray}
S_{eff}^{(2)} &=& \frac{1}{2} \int d^3q~ \delta 
A_\mu(q)M^{\mu\nu}(q) \delta
A_\nu(q) \nonumber \\
&&+ \frac{1}{2} \int d^3q~[\delta A_\mu(q) M^{\mu\nu} 
(q) \delta a_\nu(q) +
\delta a_\mu(q)M^{\mu\nu} (q) \delta A_\nu(q)] \nonumber \\
&&+ \frac{1}{2} \int d^3q~ d^3s~ \delta a_\mu(q)
 \tilde{M}^{\mu\nu}(q,s) \delta
a_\nu (s) + O(\alpha^2), 
\label{e322}
\end{eqnarray}
where

\begin{equation}
\tilde{M}^{\mu\nu} (q,s) = M^{\mu\nu} 
\delta^{(3)}(q-s) + N^{\mu\nu}(q,s).
\label{e323}
\end{equation}
Then the functional integration over $\delta a_{\mu}$ yields 

\begin{equation}
S_{eff}^{EM}[\delta A] = \frac{1}{2} \int d^3q~d^3s~ 
\delta A_\mu(q) K^{\mu\nu}
(q,s) \delta A_\nu(s)
\label{e324}
\end{equation}
where

\begin{equation}
K^{\mu\nu}(q,s)=M^{\mu\nu}(q)\delta^{(3)}(q-s)-
M^{\mu\alpha}(q)
\tilde{M}_{\alpha\beta}^{-1}(q,s)
M^{\beta\nu}(s).
\label{e325}
\end{equation}
Substituting Eq.(\ref{e323}) into(\ref{e325}), we find 

\begin{equation}
K^{\mu\nu}(q,s) = N^{\mu\nu}(q,s)+O(\alpha^2),
\label{e326}
\end{equation}
and finally

\begin{eqnarray}
S_{eff}^{EM} [\delta A] &=& \frac{1}{2} 
\int_M d^3xd^3x' \delta A_\mu(x)
N^{\mu\nu} (x,x') \delta A_\nu(x') + O(\alpha^2) 
\nonumber \\
&=& S_{CS}[\delta A_\mu] + O(\alpha^2) \nonumber \\
&=& \frac{1}{2} \int_{M_1+M_2} d^3x[\theta(-y)
\alpha_1 + \theta(y)
\alpha_2]  \epsilon ^{\mu \nu \rho} \delta A_\mu 
\partial_\nu \delta
A_\rho \nonumber \\
&&+ \frac{\alpha_1-\alpha_2}{4v}\int_{\partial M}
dtdx[(\delta A_0
-v\delta A_1)(\delta A_0 + v \delta A_1)
\nonumber
\\ &&+ (\partial_0+v\partial_1)(\delta A_0-v\delta A_1)
\frac{1}{\partial_v^2-i\epsilon}
(\partial_0+v\partial_1)(\delta A_0-v\delta A_1)]+ 
O(\alpha^2).
\label{e327}
\end{eqnarray}
It is instructive to note that, up to the first order 
in the coupling
constants $\alpha_i$'s, the FQHE in a manifold 
$M=M_1+M_2+\partial M$ 
is determined by the effective Chern-Simons action
like the integral quantum Hall effect.

The Hall current is determined by 

\begin{equation}
\left<\delta j^\mu\right> =  \frac{\delta 
S_{eff}^{EM}}{\delta A_\mu},
\label{e328}
\end{equation}
from which we find

\begin{eqnarray}
\left<\delta j^0\right>&=&\delta(y)\frac{ve^2}{h}
\left[\frac{1}{2m_1}-\frac{1}{2m_2}\right]
\frac{\partial_0+v\partial_1}{\partial_v^2-i\epsilon}
\delta E_x
+O(\alpha^2) \nonumber \\
\left<\delta j_x\right>&=&-\frac{e^2}{h}\left[\frac{1}
{2m_1}\theta(-y)+\frac{1}{2m_2}\theta(y)\right]
\delta E_y \nonumber \\
&&+\frac{ve^2}{h}\delta(y)\left[\frac{1}{2m_1}-\frac{1}
{2m_2}\right]
\frac{\partial_0+v\partial_1}{\partial_v^2-i\epsilon}
\delta E_x
+O(\alpha^2) 
\label{e329}
\\
\left<\delta j_y\right>&=&-\frac{e^2}{h}\left[\frac{1}{2m_1}
\theta(-y)+\frac{1}{2m_2}\theta(y)\right]
\delta E_x+O(\alpha^2),\nonumber
\end{eqnarray}
where $h$ is the Planck constant.
>From these results one can easily show that the total 
current in the whole
manifold $M$ is conserved:

\begin{equation}
\partial_\mu \left<\delta j^\mu\right> = 0,
\label{e330}
\end{equation}
but the current at the boundary is not:

\begin{equation}
\partial_a j^a = \frac{e^2}{h} [ \frac{1}{2m_1}-
\frac{1}{2m_2}] \delta
E_x + O(\alpha^2). 
\label{e331}
\end{equation}

To compute the explicit expression for the currents 
we need to compute
the Green's function appearing in Eq.(\ref{e329}), 
 
\begin{equation}
\frac{\partial_0 + v\partial_1}{\partial_0^2 -v^2 
\partial_1^2
-i\epsilon} f(x) = \int_{\partial M}dtdxG(x-x')f(x'),
\label{e332}
\end{equation}
which, via Fourier transformation, can be shown 
to be 

\begin{equation}
G(x) = \frac{1}{2\pi i |v|} \frac{P}{t-x/v} + \frac{1}{2|v|} \delta
(t+x/v) \epsilon(t),
\label{e333}
\end{equation}
where $P$ denotes the principal value prescription, and 
$\epsilon(t)$
is the sign function. 
For a constant external electric field $\delta E_x$, we find 

\begin{equation}
\frac{\partial_0 + v\partial_1}{\partial_v^2
-i\epsilon} \delta E_x = -\frac{x}{v} \delta E_x.
\label{e334}
\end{equation}

Using Eq.(\ref{e334}) we finally obtain the currents, 

\begin{eqnarray}
\label{e335}
\left<\delta j^0\right> &=& -\delta (y) \frac{e^2}{h}
[\frac{1}{2m_1}-\frac{1}{2m_2}]x
\delta E_x + O(\alpha^2)
\\
\left<\delta j^x\right> &=& -\frac{e^2}{h}[\frac{1}{2m_1}
\theta(-y)-\frac{1}{2m_2}
\theta(y)]\delta E_y + \delta (y)
\frac{e^2}{h}[\frac{1}{2m_1}-\frac{1}{2m_2}]x \delta E_x + 
O(\alpha^2)
\label{e336}
\\
\left<\delta j^y\right> &=& -\frac{e^2}{h}[\frac{1}{2m_1}
\theta(-y)-\frac{1}{2m_2}
\theta(y)]\delta E_x + O(\alpha^2).
\label{e337}
\end{eqnarray}
 Eq.(\ref{e337}) shows that $\left<\delta j_y\right>$, 
the current in the $y$-direction, 
is discontinuous at the boundary, $y=0$. Therefore there 
is a current 
injection to the boundary, which is responsible for the 
boundary current
in $\left<\delta j^0\right>$ and $\left<\delta j_x\right>$.
Note that the boundary current is proportional to the
variable $x$. The reason for this is that the influx 
of the charge
coming from the discontinuity of $\left<\delta j_y\right>$ 
is uniform along the
boundary, and therefore the current contribution of this 
influx will be
proportional to the length of the boundary that 
receives this influx of
charge. This is of course another realization of 
the Callan-Harvey
anomaly cancellation mechanism. 

We note that the currents (\ref{e335})$-$(\ref{e337}) are
independent of the parameter $v$, which characterizes 
the choice of
boundary conditions at $\partial M$. The reason for this
can be traced back to the Feynman Green's function
(\ref{e333}), which is dictated by the requirement of 
the existence of 
functional integral
over $\Lambda$-field at the boundary.

When we use quantum field theory to describe the 
condensed matter system
as we have been doing in this paper, what it 
describes is the situation
when the system has reached the steady state. 
Thus our result implies
that, after the system has reached the steady 
state, the Hall currents at
the boundary are $v$-independent, and given by 
Eqs.(\ref{e335})-(\ref{e337}).

However, right after the external electromagnetic 
field is applied to
the quantum Hall sample, the system is not in the 
steady state, and the
Green's function in Eq.(\ref{e329}) need not be 
the Feynman Green's
function. In this situation the boundary condition 
for the Green's
function must be supplied in accordance with the
 physical situation of
the system, and the Hall current will be
dependent on the parameter $v$.  

\section{Conclusion}

We have started from the fermion Chern-Simons theory of 
integral and fractional
quantum Hall effects defined on a manifold with 
boundary, have used only the
standard procedures of quantum field theory without 
introducing any new
assumptions, and have derived the boundary actions 
and the Hall currents
at the boundary. This method shows how the anomaly 
at the
boundary arises and also the fact that a part of the 
anomaly comes from the
boundary action (the so-called consistent anomaly) 
and the other from
the bulk action.
The structure of Hall current shows that the 
discontinuity of the Hall currents
across the boundary is responsible for the existence 
of the anomaly.
In our approach the anomaly arises naturally
after the functional integral over $\Lambda$-field. 
This is contrasted to the
approach of Ref.\cite{Balachandran} where the 
chirality constraint $((\partial_0
-\partial_1)\Lambda = 0)$ was imposed by hand.

On the common boundary of two fractional quantum Hall 
samples the edge
current along the boundary is given by
\begin{equation}
\left< \delta j_x \right> = \frac{e^2}{h}
\left( \frac{1}{2m_1} - \frac{1}{2m_2} \right)
x \delta E_x ,
\label{e41}
\end{equation}
where $\delta E_x$ is the external
electric field. This result is valid in the small 
$\alpha_i$ regime and
after the system has reached the steady state. 
During the transient period
right after the external electromagnetic field is 
turned on, the current
will be dependent on the specific boundary conditions 
characterized by
the parameter $v$. It would be interesting to find 
out the $v$-dependence
of the edge current experimentally.

Our method is based on the Carlip's method where, 
due to the gauge
variance of Chern-Simons theory at the boundary, 
the would-be gauge
degrees of freedom become dynamical at the boundary. 
This dynamical
degree of freedom is described by the $\Lambda$-field. 
Since we do not
yet know what the $\Lambda$-field does physically, 
we have integrated it
out and studied the effects through the boundary Hall current,
Eq.~(\ref{e41}). It would be of great interest 
if one could clarify
its physical implications, for example, by finding a way to
excite it by some means.

\section*{Acknowledgements}
We thank Prof. C. K. Kim for helpful discussions. 
This work was supported in
part by the Basic Science Research Institute 
Program, Korea Ministry of
Education under project No. BSRI-97-2418, BSRI-97-2425, 
BSRI-97-2427, the
Center for Theoretical Physics(SNU),  and 
the Korea Science and Engineering 
Foundation under grant No. 95-0701-04-01-3.

\end{document}